\documentclass[twocolumn]{aastex62}
\usepackage{natbib}
\usepackage{color}
\usepackage{graphicx}
\usepackage{multirow}

\newcommand{\VBLMCuno }{14.029 }
\newcommand{\IBLMCuno }{14.022 }
\newcommand{\NaEbvBLMCuno }{0.187 }
\newcommand{\HaschEbvBLMCuno }{0.198 } 
\newcommand{\avgEbvBLMCuno }{0.193 }
\newcommand{\perBLMCuno }{5.4139927 }
\newcommand{\perBLMCunoErr }{1.0e-06 }
\newcommand{\smaBLMCuno }{39.23}
\newcommand{\smaBLMCunoErr }{0.10 }
\newcommand{\daopBLMCuno }{0.00012 }
\newcommand{\daopBLMCunoErr }{1.3e-05 }
\newcommand{\fullcycleBLMCuno }{144 }

\newcommand{\aopBLMCuno }{5.41 }
\newcommand{\aopBLMCunoErr }{0.05 }
\newcommand{\eccBLMCuno }{0.0151 }
\newcommand{\eccBLMCunoErr }{0.0007 }
\newcommand{\inclBLMCuno }{80.54 }
\newcommand{\inclBLMCunoErr }{0.16 }
\newcommand{\corrgeomBLMCuno }{-0.030 \pm 0.008 }
\newcommand{\VBLMCdos }{13.714 }
\newcommand{\IBLMCdos }{13.913 }
\newcommand{\NaEbvBLMCdos }{0.088 }
\newcommand{\HaschEbvBLMCdos }{0.077 }
\newcommand{\avgEbvBLMCdos }{0.082 }
\newcommand{\perBLMCdos }{4.2707640 }
\newcommand{\perBLMCdosErr }{3.3e-07 }
\newcommand{\smaBLMCdos }{37.46 }
\newcommand{\smaBLMCdosErr }{0.11 }
\newcommand{\daopBLMCdos }{0.000145 }
\newcommand{\daopBLMCdosErr }{2.7e-06 }
\newcommand{\fullcycleBLMCdos }{118.7 }

\newcommand{\aopBLMCdos }{0.995 }
\newcommand{\aopBLMCdosErr }{0.009 }
\newcommand{\eccBLMCdos }{0.0809 }
\newcommand{\eccBLMCdosErr }{0.0008 }
\newcommand{\inclBLMCdos }{86.19 }
\newcommand{\inclBLMCdosErr }{0.3 }
\newcommand{\corrgeomBLMCdos }{0.031 \pm 0.009 }
\newcommand{\mPriBLMCuno }{13.21 }
\newcommand{\mPriBLMCunoErr }{0.08 }
\newcommand{\rPriBLMCuno }{8.48 }
\newcommand{\rPriBLMCunoErr }{0.10 }
\newcommand{\loggPriBLMCuno }{3.702 }
\newcommand{\loggPriBLMCunoErr }{0.010 }
\newcommand{\mbolPriBLMCuno }{-6.456 }
\newcommand{\mbolPriBLMCunoErr }{0.024 }

\newcommand{\logLPriBLMCuno }{4.478 }  

\newcommand{\teffPriBLMCuno }{26100 }  
\newcommand{\teffPriBLMCunoErr }{1600 }
\newcommand{\BCvPriBLMCuno }{-2.6 }
\newcommand{\BCvPriBLMCunoErr }{0.19 }
\newcommand{\mVPriBLMCuno }{15.22 }    
\newcommand{\mVPriBLMCunoErr }{0.02 }

\newcommand{\VIPriBLMCuno }{-0.02 }    
\newcommand{\VIPriBLMCunoErr }{0.03 }  
\newcommand{\VKPriBLMCuno }{-0.19 }    
\newcommand{\VKPriBLMCunoErr }{0.03 }  
\newcommand{\mSecBLMCuno }{14.44 }
\newcommand{\mSecBLMCunoErr }{0.14 }
\newcommand{\rSecBLMCuno }{12.95 }
\newcommand{\rSecBLMCunoErr }{0.05 }
\newcommand{\loggSecBLMCuno }{3.374 }  
\newcommand{\loggSecBLMCunoErr }{0.003 }
\newcommand{\mbolSecBLMCuno }{-6.962 }
\newcommand{\mbolSecBLMCunoErr }{0.008 }

\newcommand{\logLSecBLMCuno }{4.681 }  
    
\newcommand{\teffSecBLMCuno }{23750 }   
\newcommand{\teffSecBLMCunoErr }{1300 }
\newcommand{\BCvSecBLMCuno }{-2.37 }
\newcommand{\BCvSecBLMCunoErr }{0.17 }
\newcommand{\mVSecBLMCuno }{14.47 }   
\newcommand{\mVSecBLMCunoErr }{0.01 }

\newcommand{\VISecBLMCuno }{0.00 }    
\newcommand{\VISecBLMCunoErr }{0.02 } 
\newcommand{\VKSecBLMCuno }{-0.13 }   
\newcommand{\VKSecBLMCunoErr }{0.02 } 
\newcommand{\avgEbvBLMCunoErr}{0.021}
\newcommand{\vsiniPriBLMCuno }{128  $\pm$ 5} 
\newcommand{\vsiniSecBLMCuno }{115  $\pm$ 3} 
\newcommand{\vrotBFPriBLMCuno}{130 $\pm$ 5}  
\newcommand{\vrotBFSecBLMCuno}{117 $\pm$ 3}  
\newcommand{\BLMCunoDM}{18.47}        
\newcommand{\BLMCunoDMerr}{0.15}      
\newcommand{\DMcgeomBLMCuno}{18.45 $\pm$ 0.03}  
\newcommand{\DMcgeomBLMCunoDepth}{18.45 $\pm$ 0.04}
\newcommand{\mPriBLMCdos }{19.62 }
\newcommand{\mPriBLMCdosErr }{0.19 }
\newcommand{\rPriBLMCdos }{8.89 }
\newcommand{\rPriBLMCdosErr }{0.19 }
\newcommand{\loggPriBLMCdos }{3.832 }
\newcommand{\loggPriBLMCdosErr }{0.018 }

\newcommand{\mVPriBLMCdos }{14.34 }
\newcommand{\mVPriBLMCdosErr }{0.05 }

\newcommand{\VIPriBLMCdos }{-0.2 }
\newcommand{\VIPriBLMCdosErr }{0.07 }

\newcommand{\mRPriBLMCdos }{14.39 }
\newcommand{\mRPriBLMCdosErr }{0.05 }
\newcommand{\mSecBLMCdos }{19.05 }
\newcommand{\mSecBLMCdosErr }{0.14 }
\newcommand{\rSecBLMCdos }{7.92 }
\newcommand{\rSecBLMCdosErr }{0.23 }
\newcommand{\loggSecBLMCdos }{3.920 }
\newcommand{\loggSecBLMCdosErr }{0.025 }

\newcommand{\mVSecBLMCdos }{14.61 }
\newcommand{\mVSecBLMCdosErr }{0.06 }

\newcommand{\VISecBLMCdos }{-0.2 }
\newcommand{\VISecBLMCdosErr }{0.09 }

\newcommand{\mRSecBLMCdos }{14.66 }
\newcommand{\mRSecBLMCdosErr }{0.06 }
\newcommand{\wrongT}{22000}
\newcommand{\avgEbvBLMCdosErr}{0.015}
\newcommand{\vsiniPriBLMCdos }{108 $\pm$ 3} 
\newcommand{\vsiniSecBLMCdos }{105 $\pm$ 3} 
\newcommand{\vrotBFPriBLMCdos}{108 $\pm$ 3} 
\newcommand{\vrotBFSecBLMCdos}{106 $\pm$ 3} 
 
\newcommand{\DMcgeomBLMCdosDepth}{18.51 $\pm$ 0.04} 
\newcommand{\VyE}[2]{#1 $\pm$ #2 }

\accepted{for publication in ApJ}

\shorttitle{Towards early-type systems as extragalactic milestones. I}
\shortauthors{Taormina et al.}

\begin{document}

\title{Towards early-type eclipsing binaries as extragalactic milestones: I. Physical parameters of OGLE-LMC-ECL-22270 and OGLE-LMC-ECL-06782}

\correspondingauthor{M{\'o}nica Taormina}
\email{taormina@camk.edu.pl}

\author[0000-0002-1560-8620]{M{\'o}nica Taormina}
\affiliation{Centrum Astronomiczne im. Miko{\l}aja Kopernika PAN, Bartycka 18, 00-716 Warsaw, Poland}
\author{G. Pietrzy{\'n}ski}
\affiliation{Centrum Astronomiczne im. Miko{\l}aja Kopernika PAN, Bartycka 18, 00-716 Warsaw, Poland}
\author{B. Pilecki}
\affiliation{Centrum Astronomiczne im. Miko{\l}aja Kopernika PAN, Bartycka 18, 00-716 Warsaw, Poland}
\author{R.-P.  Kudritzki}
\affiliation{Universit\"atssternwarte, Ludwig Maximilian Universit\"at M\"unchen, Scheinerstr. 1, D-81679 M\"unchen, Germany}
\affiliation{Institute for Astronomy, University of Hawaii at Manoa, Honolulu, HI 96822, USA}
\author{I. B. Thompson}
\affiliation{Carnegie Observatories, 813 Santa Barbara Street, Pasadena, CA 91101-1292,  USA}
\author{D. Graczyk}
\affiliation{Centrum Astronomiczne im.  Miko{\l}aja Kopernika PAN, Rabia{\'n}ska 8, 87-100 Toru{\'n}, Poland}
\author{W. Gieren}
\affiliation{Universidad de Concepci\'{o}n, Departamento de Astronom\'{i}a, Casilla 160-C, Concepci\'{o}n, Chile}
\affiliation{Millenium Institute of Astrophysics, Santiago, Chile}
\author{N. Nardetto}
\affiliation{Laboratoire Lagrange, UMR7293, Universit{\'e} de Nice Sophia-Antipolis,  CNRS, Observatoire de la C{\^o}te d{'}Azur, Nice, France}
\author{M. G\'{o}rski}
\affiliation{Universidad de Concepci\'{o}n, Departamento de Astronom\'{i}a, Casilla 160-C, Concepci\'{o}n, Chile}
\author{K. Suchomska}
\affiliation{Centrum Astronomiczne im. Miko{\l}aja Kopernika PAN, Bartycka 18, 00-716 Warsaw, Poland}
\author{B. Zgirski}
\affiliation{Centrum Astronomiczne im. Miko{\l}aja Kopernika PAN, Bartycka 18, 00-716 Warsaw, Poland}
\author{P. Wielg\'{o}rski}
\affiliation{Centrum Astronomiczne im. Miko{\l}aja Kopernika PAN, Bartycka 18, 00-716 Warsaw, Poland}
\author{P. Karczmarek}
\affiliation{Universidad de Concepci\'{o}n, Departamento de Astronom\'{i}a, Casilla 160-C, Concepci\'{o}n, Chile}
\author{W. Narloch}
\affiliation{Universidad de Concepci\'{o}n, Departamento de Astronom\'{i}a, Casilla 160-C, Concepci\'{o}n, Chile}

\begin{abstract}
In this first paper of the series we describe our project to calibrate the distance determination method based on early-type binary systems. The final objective is to measure accurate, geometrical distances to galaxies beyond the Magellanic Clouds with a precision of $2\%$. 
We start with the analysis of two early-type systems for which we have collected all the required spectroscopic and photometric data. Apart from catalog publications, these systems have not been studied yet, and it is the first time the modeling of light and radial velocity curves is performed for them.
From the analysis we obtained precise physical parameters of the components, including the masses measured with precision of 0.6-1\% and radii with precision of 0.4-3\%.
For one system we determined the $(V-K)$ color and estimated the distance using the bolometric flux scaling method  (DM=\BLMCunoDM{}$\pm$\BLMCunoDMerr{} mag), which agrees well with our accurate determination of the distance to the LMC from late-type giants. For the same system we determined the surface brightness of individual stars using our model, and checked that it is consistent with a recent surface brightness -- color relation. 
We compared our results with evolution theory models of massive stars and found they agree in general, however, models with higher overshooting values give more consistent results. The age of the system was estimated to from 11.7 to 13.8 Myr, depending on the model.

\end{abstract}


\section{Introduction}
\label{sec:introduction}
Detached eclipsing double-lined spectroscopic binaries offer a unique opportunity to measure directly, and very accurately stellar parameters like mass, luminosity and radius \citep[e.g.][]{torres:2010}. They also provide almost purely geometrical distance determinations, with only a single empirical relation needed to be calibrated \citep{paczynski:1997,kruszewski:1999}.
The most commonly used now is the surface-brightness -- color relation (SBCR), although other options are also possible \citep{paczynski:1997}.
A precise SBCR is available for late-type stars \citep{diBenedetto:2005,pietrzynski:2019}, but the blue part of this relation is yet to be improved, as currently the precision of only $8\%$ was achieved \citep{challouf:2014}.

A lot of binary systems were observed and analyzed so far, giving precise parameters and distances, but they were mostly composed of late-type main sequence stars. However, as late-type stars are faint, such measurements are in general limited to our Galaxy. The exception are binaries composed of late-type giants, which are much brighter and in principle can be observed in other galaxies, but they are rare and hard to detect, because of their very long orbital periods (several hundred days).
They were discovered in the Large and Small Magellanic Clouds only because of a very long monitoring by the \emph{Optical Gravitational Lensing Experiment} (OGLE) survey \citep{udalski:2008}.

If we want to reach larger extra-galactic distances (beyond the Magellanic Clouds), an alternative is to observe binary systems composed of early (O and B)-type stars, which are bright enough to be observed in more distant galaxies. At the same time their orbital periods are relatively short (several days), which make them easy to discover even during a short survey.

The first to use extra-galactic early-type DEBs as distance indicators were \cite{guinan:1998} who determined the distance modulus to the Large Magellanic Cloud (LMC). Later similar systems were used to determine distances to LMC \citep{ribas:2000,fitzpatrick:2003}, M31 \citep{ribas:2005} and M33 \citep{bonanos:2006}. 
There are, however, various problems related to the distance measurement with early-type systems when a precise SBCR is not available and 
a multidimensional fit of many different parameters at the same time is necessary \citep[see][]{fitzpatrick:2003}. The differences in the distance determinations to the LMC based on the same early-type eclipsing binary $HV \, 2274$ (distance moduli in the range from $18.20$ to $18.55 \, mag$) constitute the best example of how difficult it is to measure the distance to an early-type eclipsing binary with an accuracy better than about 10 \% \citep[e.g.][]{guinan:1998, udalski:1998, nelson:2000}.

Many eclipsing binaries in galaxies beyond the Magallanic Clouds (e.g. M31, M33) were already discovered \citep{macri:2001,bonanos:2003}, of which several are detached and bright enough to serve as good distance indicators \citep{macri:2004} -- all of them being early type stars.
With future extragalactic surveys, many more early-type systems will be discovered and good quality data for fainter objects will be available with the new generation of telescopes (for example the Large Synoptic Survey Telescope). To be able to fully use these data for precise distance measurements it is important to improve the calibration of the SBCR, which is the main direct objective of our project. Our calibration will be based on the detached eclipsing early-type systems discovered in the LMC by the OGLE project \citep{udalski:2008,graczyk:2011}, for which good photometric data are available.

In the last two decades, different projects have monitored the sky in several filters with their own particular goal, but as a by-product they have left a big database of observations useful for various types of astrophysical studies.
In our region of interest, i.e., the Large Magellanic Clouds, there are three important sources of photometric data available, \emph{Exp\'{e}rience pour la Recherche d'Objets Sombres} \citep[EROS --][]{tisserand:2007}, \emph{Massive Astrophysical Compact Halo Object} \citep[MACHO --][]{alcock:1997} and OGLE projects which provide very good photometry spanning many years.
There are also three other surveys of the LMC in the near infrared, \cite{ita:2004}, \cite{macri:2015}, and \emph{VISTA near-infrared YJKs survey of the Magellanic Cloud}  \citep[VMC --][]{cioni:2011}, but their time and area coverage are quite limited with only a small number of points provided.
In our study we will use all the public photometry available for each system, complemented with new data when necessary.

We organized the paper as follows.  The project is described in section~\ref{sec:project}.
In section~\ref{sec:data} we present two of the nine systems from our sample, the data used in this study and the data reduction process. In section~\ref{sec:analysis} we explain the steps taken to derive the fundamental parameters of the systems and their components. In section~\ref{sec:results} we show full solutions for the systems and we check the consistency of our measurements with the best available SBCR. We also compare our results with the evolutionary models. 
The conclusions of our work are presented in Section~\ref{sec:summary}.

\section{Project description}  \label{sec:project}
The main goal of our project is to use early-type eclipsing binaries in the LMC to significantly improve the calibration of the blue part of the surface brightness -- color relation, which currently has a scatter of 8\% \citep{challouf:2014}.
When applied to other eclipsing binaries of this type this relation will serve to accurately and precisely measure distances to the galaxies in the Local Group and beyond. Having reliable distances to numerous galaxies with different environmental conditions (and especially -- metallicities) will be of high importance for the calibration of other standard candles, like Cepheids. This, in turn, is one of the most important steps to calibrate the cosmic distance scale. \\

\subsection{The sample}
We selected the best systems for this purpose from the OGLE catalog of 26121 eclipsing binaries in the LMC \citep{graczyk:2011}. All of them have very good quality and stable light curves with very similar eclipse depths, and are well detached with no or little proximity effects present. These properties make them the best targets for the distance determination among the all known extra-galactic early-type systems. In total we have selected nine such systems and they are all located close to the line of nodes and the center of the LMC galaxy.  They are all short-period binaries (periods from 3 to 6 days) and are bright enough to be observed in the LMC, which make them an optimal sample for the purpose of this project.

\subsection{The method}
Although in principle distances to these LMC systems could be measured directly, the expected precision from such a direct measurement is not as high as for the systems composed of late-type giants. For this reason to calibrate the SBCR we are going to use the known distance to the LMC which was measured recently with the accuracy of about 1\% \citep{pietrzynski:2019}.
We will take into account the geometry of the LMC and apply the corresponding corrections to the expected values.
Using these distances together with the radii from the modeling we can calculate the angular diameters of the stars of the system. With observed magnitudes and angular diameters we can calculate the surface brightness and obtain the SBCR for early-type stars with unprecedented precision.

\section{Observational Data}  \label{sec:data}
Here we start with the analysis of two systems, OGLE-LMC-ECL-22270 (hereafter BLMC-01) and OGLE-LMC-ECL-06782 (hereafter BLMC-02), for which we have already acquired all the planned spectroscopic data, which are necessary to obtain the reliable results. These objects were never analyzed before and nothing is known about their physical properties except a very rough spectroscopic type determination for BLMC-01.

As all the systems in our sample, these two are relatively bright (for the LMC) detached, double-lined spectroscopic binaries, with small proximity effects. These characteristics make them excellent distance indicators among extra-galactic early-type binaries analyzed so far. They have also well-defined light curves and because of their brightness it is possible to acquire for them spectroscopic data with high enough S/N.

These systems are among the most distant from the center of LMC in our sample, yet they are still close enough for the geometric corrections (due to the inclination of the LMC disk) to be relatively small. For example, using the geometric model of this galaxy of \citeauthor{vandermarel2014} (\citeyear{vandermarel2014}; PMs+Young) we obtain corrections: $\corrgeomBLMCuno$ mag for BLMC-01 and $\corrgeomBLMCdos$ mag for BLMC-02.

\begin{figure*}
    \begin{center}
        \includegraphics[width=\textwidth]{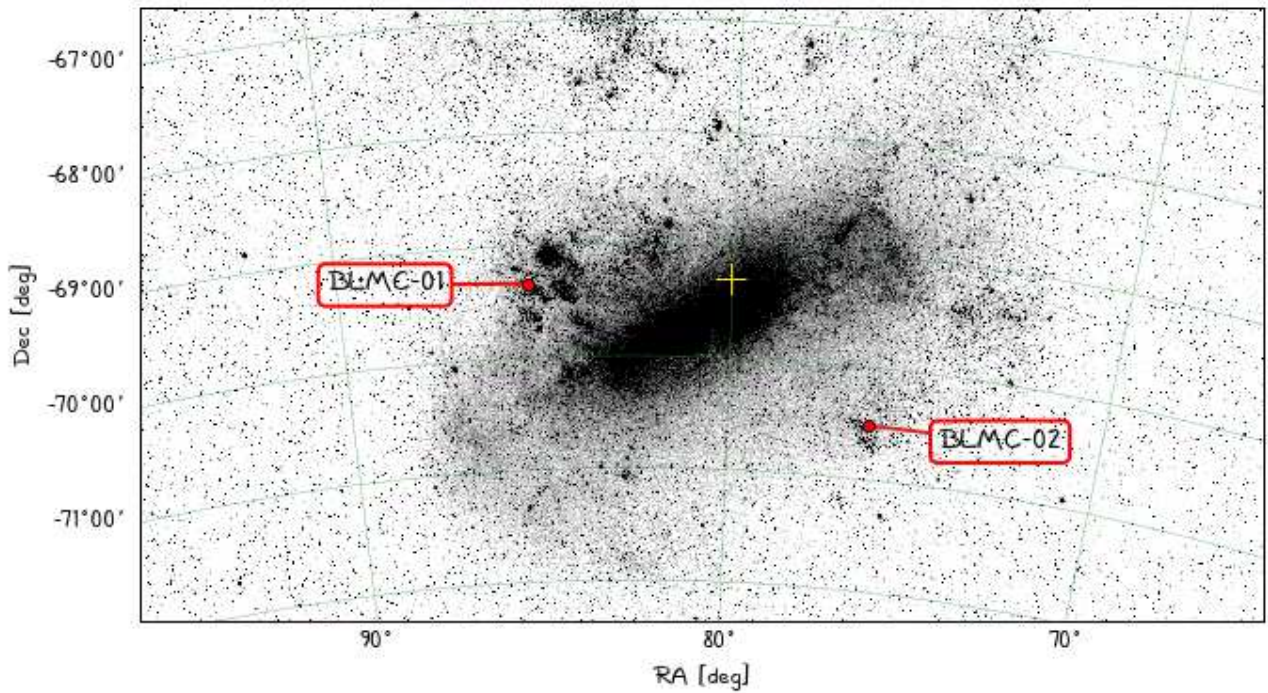} 
        \end{center}
    \caption{Position of the OGLE-LMC-ECL-22270 (BLMC-01) and OGLE-LMC-ECL-06782 (BLMC-02) systems in the sky. The center of LMC (from the model of \citealt{vandermarel2014}) is marked with a cross.}
    \label{skyposition}
\end{figure*}

BLMC-01, is a detached eclipsing binary (DEB) system located in the north-west (closer to us) part of the LMC, in the 30 Doradus star-forming region (see Fig.~\ref{skyposition}), which is known to have the highest internal extinction levels in the LMC. The spectral type of BLMC-01 was determined as B1 III by \cite{muraveva:2014}, while \cite{evans:2015} found that it is a double-lined spectroscopic binary composed of two early type stars. The orbital period of BLMC-01 is 5.41 days and its V-band brightness at maximum is \VBLMCuno mag.

BLMC-02 is located on the opposite side of the galaxy center, in its south-east part -- see Fig.~\ref{skyposition}. Its orbital period is 4.27 days and its V-band brightness at maximum is \VBLMCdos mag.
General information about the systems is given in Table~\ref{info}.

\begin{table*}
    \caption{General information from the literature. }
    \begin{center}
    \begin{tabular}{|c |c | c|c || c || c|c |c|}
        \hline
         \multirow{2}{*}{Our ID} &  \multirow{2}{*}{OGLE ID}  &  \multicolumn{2}{c||}{Coordinates} & Orbital Period &\multicolumn{2}{c|}{Brightness} & Color \\
        \cline{3-8}
                                 &   & $\alpha_{2000}$     & $\delta_{2000}$    &     [days]         & I        & V    & $(V-I)$\\
        \hline\hline
        BLMC-01 & OGLE-LMC-ECL-22270 & 05:40:47.31  & -69:20:28.3 &  5.414011  & \IBLMCuno & \VBLMCuno & 0.007   \\
        BLMC-02 & OGLE-LMC-ECL-06782 & 05:04:59.05  & -70:31:13.9 &  4.270776  & \IBLMCdos & \VBLMCdos & -0.199  \\
        \hline 
    \end{tabular}
    \end{center}
    \label{info}
\end{table*}

\subsection{Photometry}
We have collected all the photometric data available for the two systems up to now. For both, the most important data (in the sense of the coverage range, data quality and the use of standard filters) come from the OGLE project.
The OGLE observations were carried out during the third and fourth phase of this microlensing survey with the 1.3-m Warsaw telescope at the Las Campanas Observatory. In our analysis we make use of their measurements taken in the I and V-bands of the Johnsons-Cousins system. 

For the BLMC-01 system in total we collected 821 observations in the I-band and 116 data points in the V-band from OGLE-III and OGLE-IV, while for BLMC-02 we collected 1004 and 187 data points for the I and V-band, respectively.

The BLMC-01 system was also observed by the EROS-2 survey, where its ID number is lm0031l22987.
The survey was carried out with the Marly 1-m telescope at ESO, La Silla from 1996 July to 2003 February. Observations were performed in two (blue and red) wide passbands with central wavelengths at 490 nm ($B_{EROS}$) and 670 nm ($R_{EROS}$), respectively. 
The filters used are not standard, and for this reason in our analysis we have only used the $R_{EROS}$ band which is centered close to the $I_{C}$ standard band and can be transformed to Johnson-Cousins standard system with a precision of $\sim 0.1$ $mag$, using the simple relation from \cite{tisserand:2007}: $R_{EROS} = I_{C}$.
We used 421 measurements from this survey, which allowed us to obtain a greater observational coverage in time for the I-band light curve -- in total 1242 data points spanning over 17 years for EROS-2 and OGLE data, together. 

This system is also located in one of the fields of the VMC survey, where its ID is J054047.30-692028.25. These survey observations were taken using the VISTA 4.1-meter telescope located at the Paranal Observatory in Chile during one of the ESO Public Surveys. In the photometric catalog of the VMC survey there are 14 measurements in the $K_S$ band.  Fortunately both eclipses were covered and we could use these observations in the light curve modeling.

Unfortunately BLMC-02 was not observed by the EROS and VMC projects, but additional observations from the MACHO survey are available. MACHO observations were carried out with the 1.27-meter telescope at Mount Stromlo Observatory,
Australia, in two photometric bands (red and blue), which can be transformed to the standard $R_C$ an $V_J$ bands using relations from \citet{faccioli:2007} with the accuracy of 0.035 mag \citep{alcock:1999}. The star ID in the MACHO database is 23.3908.17.

\begin{figure}
    \begin{center}
        \includegraphics[width=0.49\textwidth]{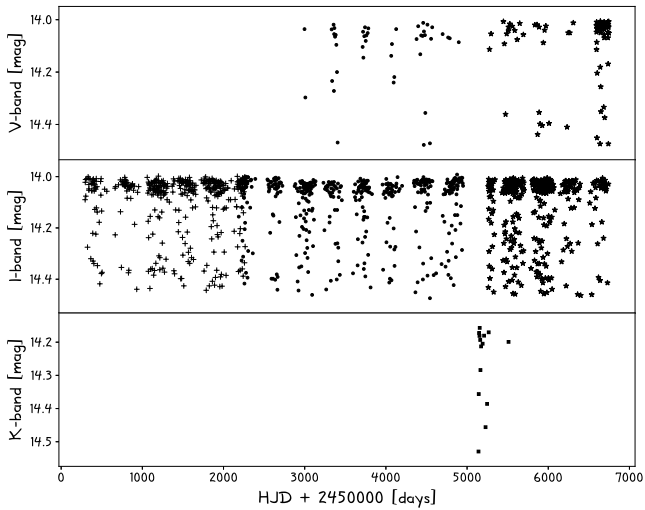}
    \end{center}
    \caption{All the photometric data collected for BLMC-01. In the upper panel we show the V-band photometry from the OGLE3 (circles) and OGLE4 (stars), in the middle panel the I-band data from EROS-2 (crosses), OGLE3 and OGLE4. In the bottom panel the K-band measurements from VMC (squares) are presented.}
    \label{photdata29}
\end{figure}

\begin{figure}
    \begin{center}
        \includegraphics[width=0.49\textwidth]{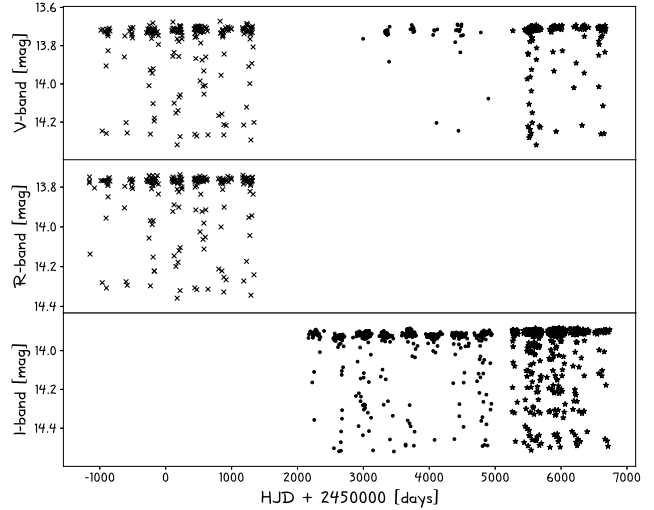}
    \end{center}
    \caption{All the photometric data collected for BLMC-02. In the upper panel we show the V-band photometry from the OGLE3 (circles), OGLE4 (stars) and MACHO ($\times$ symbols), and in the middle panel the MACHO R-band data. In the bottom panel the I-band photometry from OGLE3 and OGLE4 are presented.}
    \label{photdata22}
\end{figure}

In Table~\ref{points} for both systems we show the numbers of the measurements for all the surveys described above and in Fig.~\ref{photdata29} and \ref{photdata22} we present a graphical summary of all the photometric data used in the analysis.

\begin{table*}
    \caption{Photometric data }
    \begin{center}
    \begin{tabular}{|c || c|c || c|c || c || c|c ||c|}
        \hline
         \multirow{2}{*}{Our ID} &
         \multicolumn{2}{c||}{OGLE-III} &  \multicolumn{2}{c||}{OGLE-IV} &
         EROS-2 & \multicolumn{2}{c||}{MACHO}  & VMC \\
        \cline{2-9}
                &  I  &  V  & I  &  V  & $R_{eros}$ & R   &  V  &  Ks \\
        \hline\hline
        BLMC-01  & 437 & 384 & 45 & 71  &   421      & --- & --- & 14 \\
        BLMC-02  & 412 & 592 & 39 & 148 &   ---      & 196 & 201 & --- \\
        \hline
    \end{tabular}
    \end{center}
    \label{points}
\end{table*}

\subsection{Spectroscopy}
Although our systems are relatively bright for objects located in the LMC, they are still not easy targets for spectroscopic observations and one needs world's best telescopes to obtain good quality data with reasonable exposure times. We have carried out such observations at two sites in Chile.

The majority of our high-resolution optical spectra were acquired with the Very Large Telescope (VLT) and the Ultraviolet and Visual Echelle Spectrograph \citep[UVES --][]{dekker:2000} mounted at the Nasmyth B focus of UT2 Kueyen at Paranal Observatory.
Observations were taken with the UVES standard configuration DIC1 390+564 which provided a coverage of two wavelength ranges 3260-4520 \AA{} and 4580-6690 \AA{}. We used a slit of 0.7 arcsec for the red part and 0.9 arcsec for the blue part of the spectra, obtaining a spectral resolution of R $\sim$ 50000. The exposure times per spectrum ranged from 20 to 30 minutes. 

Additional spectra were obtained with the red and blue arms of the MIKE spectrograph \citep{bernstein:2003} on the 6.5m Magellan Clay Telescope at the Las Campanas Observatory. All the observations were taken with a 0.7 arcsec slit at a resolution of about 40000 for the red side and about 50000 for the blue side. The spectra were taken with the integration times from 10 to 30 minutes.

A total of fourteen high-resolution spectra (10 UVES + 4 MIKE) were obtained for BLMC-01 system and twelve (8 UVES + 4 MIKE) for BLMC-02. Typical signal to noise ratios near $4000$ \r{A}  are about 50-60 for UVES spectra and about 30-40 for the MIKE spectra.

For the reduction of the UVES spectroscopic data we used {\tt UVES Workflow version 5.8.2} running in the {\tt ESO Reflex} environment {\tt version 2.8.5} \citep{freudling:2013}. This pipeline automatically follows the basic reduction echelle steps and returns the extracted one-dimensional spectrum (blue and red parts). 
MIKE data were reduced using Daniel Kelson's pipeline available at the Carnegie Observatories software Repository\footnote{http://code.obs.carnegiescience.edu/}.

\section{Analysis} \label{sec:analysis}
The final modeling of the presented systems is done using the Wilson-Devinney code (hereafter {\tt WD}, \citealt{wilson:1971}, \citealt{vanHamme:2007}) with the {\tt PHOEBE} graphical interface \citep{prsa:2005}. The WD code is widely used and very well-tested. It is also the best tool for the modeling of non-spherical stars, i.e., stars that are close enough that tidal forces are deforming the shape of the stars and/or those affected by fast rotation. It comes at a cost however -- this code is very slow, especially for eccentric systems.

Both systems analyzed here are eccentric and for both of them the apsidal motion was detected. For this reason we decided to first derive approximate parameters using faster and less accurate tools and then use these results as initial parameters for the {\tt WD} code to improve the solution. We started with the derivation of the orbital parameters using the {\tt RaveSpan} (Radial Velocity and Spectrum Analyzer) {\tt code} presented in \citet{pilecki:2017} (see also: \citealt{pilecki:2018a}), and then we obtained the preliminary light curve models using a modified version of {\tt JKTEBOP} code \citep{southworth:2007}, which allowed us also to estimate the apsidal motion.  Then the final modeling of the light and radial velocity curves was performed, followed by the determination of extinction and temperature. Eventually, the absolute parameters were determined.

\subsection{Derivation of orbital parameters} \label{sec:pre_orbpars}
\begin{figure}
    \begin{center}
        \includegraphics[width=0.47\textwidth]{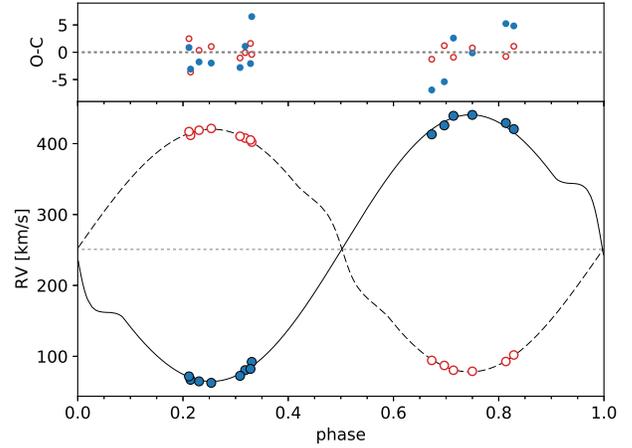} \\
    \end{center}
    \caption{The orbital solution from the {\tt WD} code and the radial velocity measurements from {\tt Ravespan} for BLMC-01. Filled circles -- primary component, open circles -- secondary component. Residuals are shown in the upper panel. The orbital RVs presented here are available as the data behind the figure.}    
    \label{rvc01}
\end{figure}

\begin{figure}
    \begin{center}
        \includegraphics[width=0.47\textwidth]{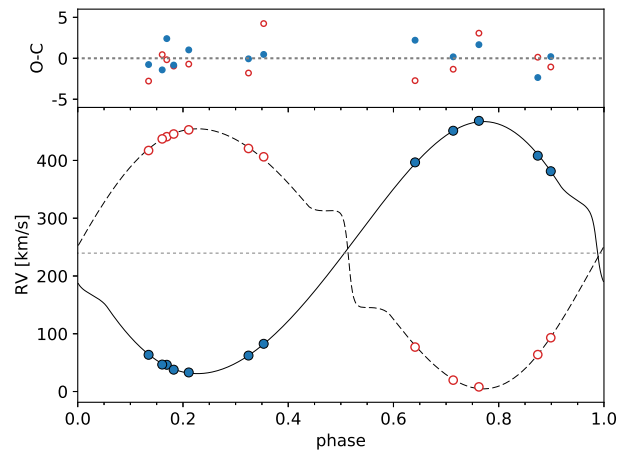} \\
    \end{center}
    \caption{The orbital solution and the radial velocity measurements for BLMC-02. The same symbols as in Fig.~\ref{rvc01} are used. The orbital RVs shown here are available as the data behind the figure.}
    \label{rvc02}
\end{figure}

To obtain the orbital solution, first radial velocities have to be determined.
For that we used the {\tt RaveSpan} code which is a user-friendly graphical Python application that allows extracting velocities of the components using such methods as cross-correlation function (CCF), TODCOR, and Broadening Function (BF).
We applied the BF technique \citep{rucinski:1999} because it gives narrower and stronger profiles than the commonly used CCF, which is very important for early-type stars with strong rotational broadening.
As spectrum templates we used spectra of standard stars from the ESO Data Archive. For each system we have selected various templates with the spectral type similar to that estimated for the components.   
The radial velocity measurements for BLMC-01 are shown in Fig.~\ref{rvc01} and for BLMC-02 in Fig.~\ref{rvc02}.

The next step was to fit the obtained radial velocity curves (RVC). In the model we took into account the following orbital elements of the spectroscopic binary: $P$ (the orbital period), $K_{A}$ and $K_{B}$ (the orbital semi-amplitudes), $\gamma$ (the velocity of the center-of-mass of the system), $T_0$ (the reference time), $e$ (the orbital eccentricity) and $\omega$ (the argument of periastron passage of the primary). Assuming the Keplerian motion of the point-like sources the radial velocities of the components depend on these parameters in a following way: 
\[ v_{i} = \gamma + K_{i} \, e \, cos w + K_{i} \, e \, cos(\nu + w), \]
where $\nu$ is the true anomaly. 
We fitted this relation to the observed RVC of each component of the system and obtained the preliminary orbital solution. 

The minimum masses of the components (expressed in the solar mass units) can be calculated as follows: 
\[  m_{A,B} \, sin^{3} i = 1.036 \times 10^{-7} (K_{A} + K_{B})^{2} K_{B,A} \, P \, (1-e^{2})^{3/2},  \]

where $i$ is the orbital inclination and $P$ and $K_{i}$ are expressed in days and $km/s$, respectively. Because $sin^{3} i \le 1$, what we can measure here is just the \emph{minimum mass}. The mass ratio is inversely proportional to the ratio of the semi-amplitudes, i.e. $q = m_{B}/m_{A} = K_{A}/K_{B}$, and is independent of inclination.
The orbital parameters from this analysis for both systems are presented in Table~\ref{orbparams}.

\begin{table*}
\begin{center}
\caption{Orbital parameters and rotation from the {\tt Ravespan} code.}
\begin{tabular}{|l|cc||cc|}
\cline{1-5}
                                       &  \multicolumn{2}{c||}{  BLMC-01}         &   \multicolumn{2}{c|}{ BLMC-02} \\
\hline
Parameter                              & Primary & Secondary                       & Primary & Secondary  \\
\hline\hline
orbital period,  $P$ [days]              & \multicolumn{2}{c||}{\perBLMCuno }          & \multicolumn{2}{c|}{ \perBLMCdos} \\
eccentricity, $e$                        & \multicolumn{2}{c||}{ 0.016 $\pm$ 0.005 } & \multicolumn{2}{c|}{ 0.078 $\pm$ 0.004 }     \\
argument of periastron, $\omega$ [rad]         & \multicolumn{2}{c||}{6.0 $\pm$ 0.3 } & \multicolumn{2}{c|}{ 1.61 $\pm$ 0.07 }     \\
$m\sin^3 i$ [$M_\odot$]       & 12.38 $\pm$ 0.16 & 13.50 $\pm$ 0.24       & 19.40 $\pm$ 0.14 & 18.86 $\pm$ 0.14\\ 
Orbital semiamplitud, $K$ [km~s$^{-1}$]        &  187.1 $\pm$ 1.5  & 171.6 $\pm$ 0.8  &    218.7 $\pm$ 0.7 & 224.9 $\pm$ 0.7  \\ 
$a\sin i$ [$R_\odot$]            & 20.03 $\pm$ 0.16 & 18.40 $\pm$ 0.09       & 18.41 $\pm$ 0.06  & 18.93 $\pm$ 0.07        \\
systemic velocity, $\gamma$ [km~s$^{-1}$]                        & \multicolumn{2}{c||}{ 250.9 $\pm$ 0.6 } & \multicolumn{2}{c|}{239.6 $\pm$ 0.4 }     \\
mass ratio, $q = m_2/m_1$                                    & \multicolumn{2}{c||}{1.091 $\pm$ 0.010 } & \multicolumn{2}{c|}{0.972 $\pm$ 0.004 }     \\
$v_{rot} \, \sin i$ [km~s$^{-1}$]   &  \vsiniPriBLMCuno       & \vsiniSecBLMCuno  & \vsiniPriBLMCdos & \vsiniSecBLMCdos  \\
rms [km~s$^{-1}$]                   &  3.91 &  1.49    &   1.39  & 2.02  \\
\hline
\end{tabular}
\end{center}
\tablecomments{ Formal errors from the fitting procedure are given. Orbital periods were fixed at photometric values. Argument of periastron for a mean epoch of spectroscopic observations of a system is given. }
\label{orbparams}
\end{table*}

Early-type stars are known for their fast rotation, which is responsible for the strong broadening of the absorption lines in the spectra. In the broadening function method, that we used for the radial velocity determination, the rotation velocity of the components is also measured. We have determined these values for various spectra taken close to the orbital quadratures, where the profiles are well separated, and measured the average (projected) rotation velocity of each component of the system ($v_{1,2} \sin i$). They are given in Table~\ref{orbparams}.

\subsection{Preliminary light curve model}
We performed a preliminary light curve modeling of the system using a Python wrapper for the {\tt JKTEBOP} code \citep{southworth:2007}, which is based on the {\tt EBOP} code \citep{popper:1981}. 
In {\tt JKTEBOP} the stars are represented as biaxial spheroids but the spherical approximation is used for the calculation of the light lost during the eclipse. Using our wrapper the determination of apsidal motion is also possible with this code. 

Because of small deformations present in the light curve that come from the tidal distortions of the components, {\tt JKTEBOP} could be used only to obtain a first model approximation. It is however extremely fast when compared to the {\tt WD} code that we use later to improve the binary models, especially for eccentric systems. It let us scan relatively quickly a large parameter space and look for a solution that corresponds to a global minimum. A common problem in using differential corrections implemented in the {\tt WD} code is that the solutions tend to stack at any local minimum that is first found.

We modeled all the photometric data sets simultaneously, which in case of BLMC-01 are I, V and K bands, and in case of BLMC-02 -- I, V and R.
All the photometric datasets were first detrended and cleaned of significantly (higher than 5-sigma) outlying points.

In the model we fitted: the orbital period, the reference time, the eccentricity and argument of periastron of the primary represented in the code as $e \sin \omega$ and $e \cos \omega$, relative stellar radii, the inclination, and the surface brightness ratio and third light in each band. In this model the component temperature is only used to calculate the limb darkening coefficients. At the beginning the temperature was estimated from the dereddened system color (see Section~\ref{subsec_extinction}), and later updated according to the color of the components once a stable solution was found.

To find the best model we started from a scan of a wide parameter space containing all possible solutions for this type of binaries. Then we used the Markov Chain Monte Carlo (MCMC) method starting from all the local minima found, looking for a solution with the lowest $\chi^2$ value.

After the closer examination of the light curves residuals we found an evidence of the apsidal motion and also added the change of the position of the periastron ($\dot{\omega}$) to the list of fitted parameters.
We repeated the modeling with the apsidal motion taken into account. The best solution from this analysis was then used as a starting point for the final analysis with a more complex modeling.

\subsection{Apsidal motion}
In the analysis of variable stars and in the modeling of the binary systems in particular, it is very important to have a long baseline, which improves the determination of the period and let us also study additional secular effects.
In our case we have almost 20-year long coverage, which for orbital periods of 4-5 days, can give a very high precision in the determination of the time-related parameters, like period and times of minima. This in turn let us detect and measure the precession of the line of apsides, which is a very common phenomena in massive binary systems \citep{garcia:2014,pablo:2015,rauw:2016,zasche:2019}.

Because of low eccentricity the effect for BLMC-01 could be barely noticed looking at a light curve from one data set, although the timespan of the observation was long (7.5 years in case of OGLE-III). During the analysis of the light curves we have noticed however that in data sets from different surveys the position of eclipses was slightly different. We have identified this change as an evidence of apsidal motion in this system. The apsidal motion of BLMC-02, which is more eccentric, is much stronger and affects significantly the eclipses positions even within the time span of one dataset.

We have used a very simple method to measure this effect in the {\tt JKTEBOP} code. We have added just one parameter to the model, a derivative of the longitude of periastron ($\dot{\omega}$), which is used to calculate a constant $\omega$ for every light curve. Then for every survey we have calculated an average time of observation $<T_{i}>$.
The longitude of periastron for given light curve is then calculated as:

$$ \omega = \omega_{0} + \dot{\omega} \times (<T_{i}> - T_{0}), $$

where $\omega_{0} = \omega(T_{0})$.

As a result, we have a model for each light curve that shares all but one parameter, with the only difference of $\omega$, which influences the position of the eclipses.
This way we obtain just one solution that fits well all the light curves without the need of fitting different models for different epochs and then averaging them. 
The values of $\omega_0$ and $\dot{\omega}$ from the best model were used as input values for the WD model. The final values of the apsidal motion for each system are given in the next section.

\subsection{Final modeling of light and radial velocity curves}
\label{subsec:finmod}
For BLMC-01 the apsidal motion is weak, but the proximity effects (deformation of the stars) are significant; for BLMC-02 we have an opposite situation -- strong apsidal motion and practically insignificant proximity effects. Because neither of these phenomena are natively included in the {\tt JKTEBOP} we decided to model these systems using the widely-used and well-tested {\tt WD} code, which takes these effects into account.

This code works much slower, especially in case of eccentric systems, which makes it very important to have an approximate solution and a well studied parameter space to avoid the traps of local minima in the modeling process. Our solution from the {\tt JKTEBOP} code meets these conditions.
As initial input parameters for the {\tt WD} code we have used the parameters of the model with the lowest global $\chi^2$ found during the Monte Carlo analysis using the {\tt JKTEBOP} code. 

We modeled photometric and spectroscopic data simultaneously, so this time the model included more parameters: the orbital period (P), the reference time ($T_0$), the semi-major axis ($a$), the mass ratio ($q$), the systemic velocity ($\gamma$), the eccentricity ($e$) and argument of periastron of the primary (hotter) component ($\omega$), the first derivative of the argument of periastron ($\dot{\omega}$), the inclination ($i$), surface potentials ($\Omega_1$ and $\Omega_2$), the third light ($l3$) and the temperature of a less luminous component ($T_{eff,i}$). The temperature of a more luminous component was estimated from its dereddened (preferentially V-K) color using a transformation of \citet{worthey:2011}.  At the beginning the color from the preliminary solution was used. Later it was being updated according to the most recent solution if the change was higher than $\sim20\%$ of the estimated error. Please also note that the used color--temperature transformation was obtained for solar metallicity stars, while the metallicity of the LMC is about half the solar.

To look for the solution, we first used the differential correction method implemented in the code, which we repeated until the convergence was obtained. Then we ran a Monte Carlo simulation, calculating about 50000 models to find the $\chi^2$ minimum and estimate the errors in a reliable way. Eventually, we took the model with the lowest $\chi^2$ value as the final solution.

\begin{figure}
    \begin{center}
        \includegraphics[width=0.50\textwidth]{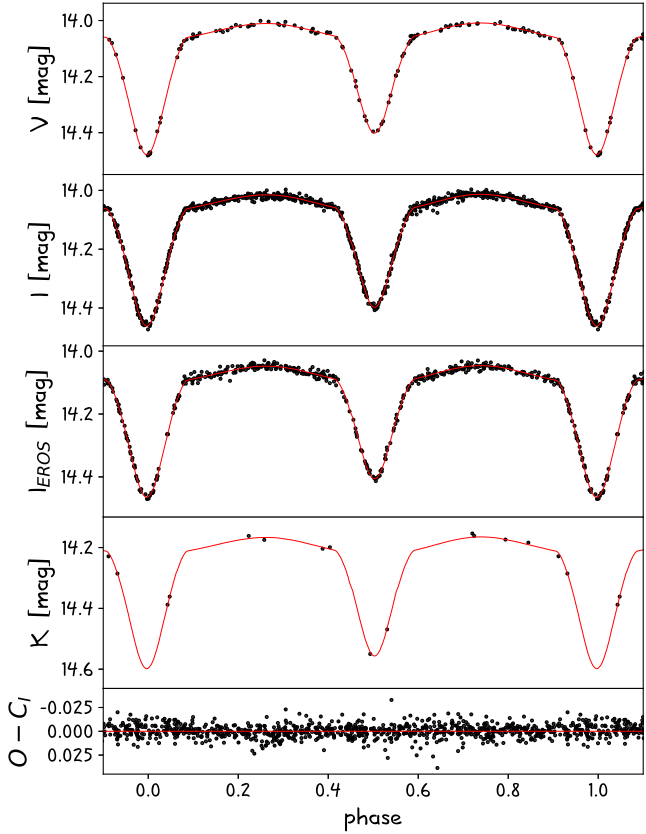} \\
    \end{center}
    \caption{Wilson-Devinney model of the I, V and K-band light curves for BLMC-01. Red line -- a model with $\omega$ fixed at the reference time for the system. Black points -- data corrected for apsidal motion. In the bottom panels I-band residual light curve is shown.
    }
    \label{LCfit_blmc29}
\end{figure}

\begin{figure}
    \begin{center}
        \includegraphics[width=0.50\textwidth]{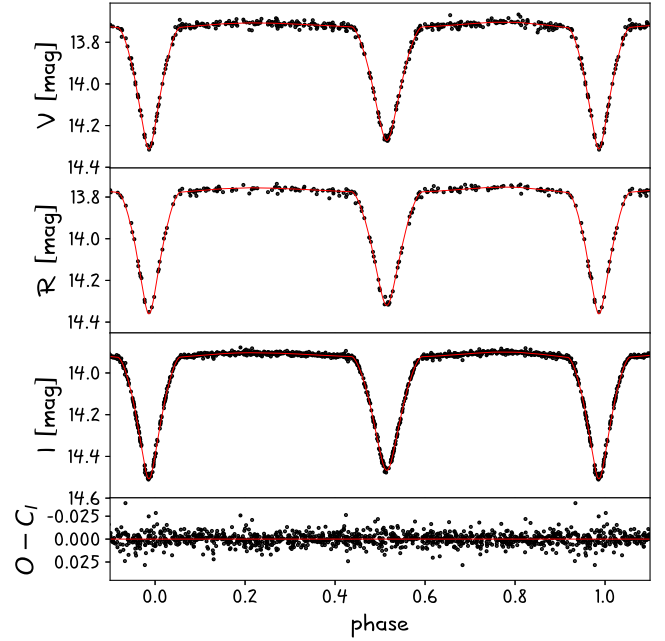} \\ 
    \end{center}
    \caption{Wilson-Devinney model of the I, V and K-band light curves for BLMC-02. For description see Fig.~\ref{LCfit_blmc29}. }
    \label{LCfit_blmc22}
\end{figure}

The final light curve models for BLMC-01 and BLMC-02 are shown in Fig.~\ref{LCfit_blmc29} and Fig.~\ref{LCfit_blmc22}, respectively. It is not easy to present the light curve for a binary system with non-zero apsidal motion. To solve this problem we corrected the observational data for $\dot{\omega}$ from the model (black points) and plotted the model light curve for a fixed epoch ($T_0$).
The correction was performed by applying to every measured magnitude a shift between the model magnitude at $T_0$ and the model magnitude at the time of observation.

\begin{figure}
    \begin{center}
        \includegraphics[width=0.50\textwidth]{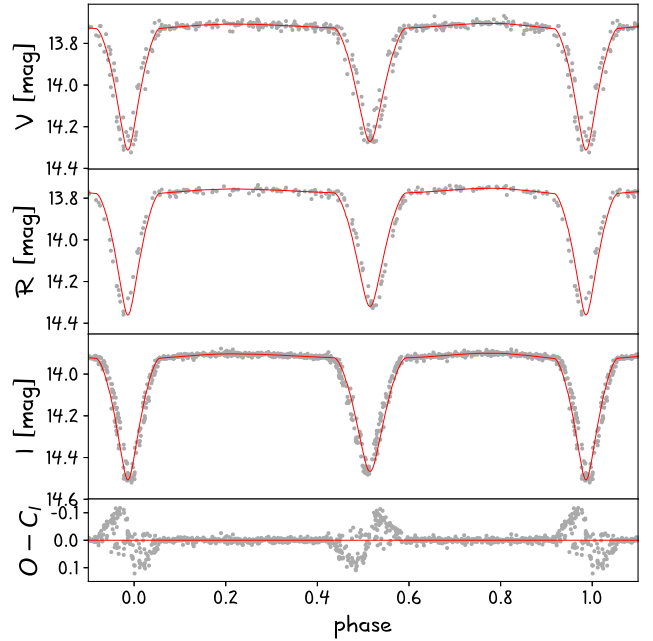} \\ 
    \end{center}
    \caption{Same as Fig.~\ref{LCfit_blmc22}, but without taking into account the apsidal motion.
    }
    \label{LCfit_blmc22am}
\end{figure}

To demonstrate the effect of apsidal motion on the light curve, in Fig.~\ref{LCfit_blmc22am} we show the same model and data but without applying the correction.
The difference is especially noticeable at eclipses, where the observations are shifted to the left or right of the model, in the opposite direction for the primary and secondary eclipses. In the bottom panel of the figure (residuals) one can see the effect is very significant.
For BLMC-01 the effect is much weaker, but also visible and detected with high confidence (9-$\sigma$ detection) using all sets of data. The full apsidal motion cycle takes \fullcycleBLMCuno{} years for BLMC-01 and \fullcycleBLMCdos{} years for BLMC-02.

For BLMC-01 we have detected a negligible third light in the OGLE I-band light curve ($<0.5\%$) and a very small negative third light in the OGLE V-band ($\sim 1\%$), which may result from the uncertainties in the flux calibration process. Only the EROS light curve was significantly affected, with the additional flux of the level of almost $6\%$. The additional/missing flux was taken into account in calculating the components magnitudes in each band. Because of low number of points, the third light for K-band light curve was not fitted. For BLMC-02 no third light was detected in any band.

The best fits to the radial velocity curves from the {\tt WD} code for the BLMC-01 and BLMC-02 systems are shown in Fig.~\ref{rvc01} and Fig.~\ref{rvc02}, respectively.

\subsection{Extinction}
\label{subsec_extinction}
The extinction in the direction of our system was estimated in two independent ways. First, we used the reddening map for the Magellanic Clouds obtained from the analysis of red clump stars \citep{haschke:2011}. We calculated the color excess using the equation:

$$ E(B-V) = \frac{E(V-I)}{1.3} + 0.04 \; mag,$$

where $E(V-I)$ is the color excess from the reddening map and the denominator of 1.3 is adopted from \citet{bessell:1998}.
As the zero-point of this map is not well-defined to obtain absolute values of $E(B-V)$ we used a shift of 0.04 mag, which results from the determination of the reddening-free mean Red Clump color $(V-I)^{RC}_0$ given by \citet{pawlak:2016}. 

We obtained the reddening towards BLMC-01 $E(B-V) = \HaschEbvBLMCuno \pm 0.029$ mag, which is higher than an average for LMC, but it was expected as the object is located in the 30 Dor region.
The reddening to the BLMC-02 system was estimated in the same way to be $\HaschEbvBLMCdos \pm 0.023$ mag.

\begin{figure*}
    \begin{center}
        \includegraphics[width=\textwidth]{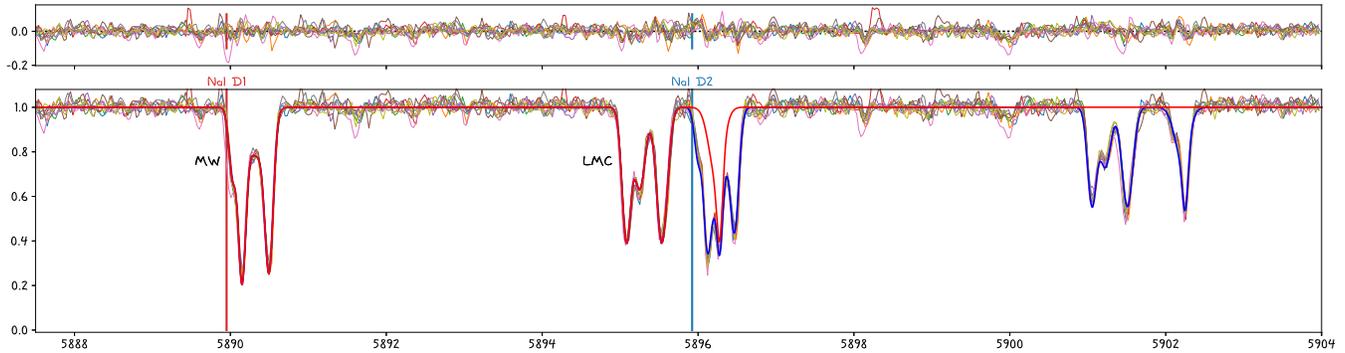} 
    \end{center}
\caption{Absorption lines of the sodium doublet and a multi-gaussian fit for BLMC-01. Red line shows a fit to the Na D1 line only and the blue one to the Na D1 and D2 lines together. Multiple components in the Galaxy ($\sim$ 5890 \AA) and in the LMC (5895 \AA) were detected. Note that at 5896.4 \AA{} LMC D1 component is superimposed on the MW D2 component. No other significant line is present in the spectra.}
\label{sodium_blmc29}
\end{figure*} 

\begin{figure*}
    \begin{center}
        \includegraphics[width=\textwidth]{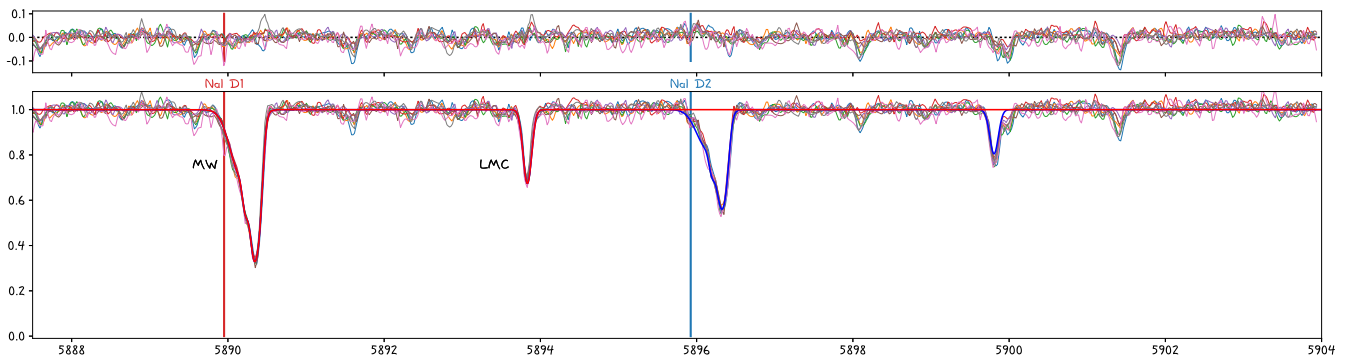} 
    \end{center}
\caption{Same as Fig.~\ref{sodium_blmc22}, but Na D1 and D2 lines for BLMC-02 are shown. The LMC component at 5894 \AA{} is much weaker than for BLMC-02 and the overall structure is much simpler. }
\label{sodium_blmc22}
\end{figure*} 

The second method used to estimate the reddening was measuring the equivalent widths of Na I sodium doublet lines (around 5893 \AA) and using the calibration of \cite{munari:1997}. This method works well for early-type stars because virtually no other lines are present in the vicinity, making the measurement more robust compared to late-type stars with many metal lines.
Nevertheless the sodium line profiles can be quite complex, as different components with different wavelength shifts may contribute to their overall shape.

To estimate the contribution of each component we fitted multiple Gaussians to the Na I profiles. The fit was performed simultaneously to several best quality spectra from different orbital phases. The final value of E(B-V) was then calculated as a sum of reddenings of individual components.

In Fig.~\ref{sodium_blmc29} and Fig.~\ref{sodium_blmc22} we show the spectra in the region of the Na I lines and the profile fits for each system. As can be seen, stellar lines (that should vary with the orbital phase) are not present in this region.
In the case of BLMC-01 there are two main Na I components related to the Galaxy and a complex structure of four components related to the LMC, where one component of D1 is mixed with the Galactic components of the D2 line. In case of BLMC-02, which is positioned on the other side of the LMC and farther from the center, there is only one weak component related to that galaxy and the majority of the extinction comes from the Milky Way (MW).

Using the method described above we determined the reddening E(B-V) of $\NaEbvBLMCuno \pm 0.031$ mag for BLMC-01 and $\NaEbvBLMCdos \pm 0.019$ mag for BLMC-02, which are consistent with the values obtained from the reddening map. Eventually we used the average of both determinations as final values, i.e., E(B-V) of $\avgEbvBLMCuno \pm 0.021$ mag for BLMC-01 and $\avgEbvBLMCdos \pm 0.015$ mag for BLMC-02. We then used these values to deredden all the colors and to calculate absorption (A$_{\lambda}$) in each band using the standard reddening law relations from \citet{cardelli:1989} and the ratio of total-to-selective extinction $R_V = 3.1$. 

For the BLMC-02 system, the temperature estimated from the dereddened (V-I) color is, however, very low ($\sim$ \wrongT ) which is not consistent with the observed spectral line features. Moreover, it also leads to an unrealistically short distance to the LMC.
The problem may partly come from the uncertain  $T - (V-I) $ calibration \citep{worthey:2011}, which is not well defined for hot stars, or with the calibration of the photometry, as a difference of 0.03 mag in $V-I$ color translates to a difference of about 4000 K. We also cannot exclude, that for some reason the reddening determinations for BLMC-02 are wrong, although two independent methods gave a similar result.

For BLMC-01 we obtain a good consistency between the temperature from the dereddened V-K color \citep{worthey:2011} and the observed spectral features. The determined value of E(B-V) also leads to a distance to LMC consistent with the accurate value obtained by \citet{pietrzynski:2019}.

We checked that for BLMC-02 to obtain a consistent distance ($DM \sim 18.5$ mag) and spectral type (O-type) and assuming our reddening estimation is correct, the temperature of the stars should be around 34000-35000 K.  
If we, however, assume that the photometry and temperature-color calibration are correct, to obtain a consistent distance we would need the reddening of $E(B-V) = 0.18$ mag, which would imply a temperature of about $42000$ K. Such high values are rather excluded, so for the moment we accept the first option as the more probable one. In the final modeling we used $T_1=35000$ K as the temperature of the primary (e.g. for calculation of the limb darkening coefficients).

In the future we plan to determine the temperature and E(B-V) for BLMC-02 directly from the spectra. This detailed spectral analysis will be published in a follow-up paper.

\section{Results}\label{sec:results}
In the previous section we showed how the orbital solutions and the light curve models for our systems were obtained and also how the reddening toward the systems was estimated. Using all this information we derived the physical parameters of the components. They are presented in Table~\ref{properties}. For both systems hotter components were chosen as primaries, but in case of BLMC-02 the primary is also the more massive one. Hotter components were chosen according to the surface brightness ratios ($j_{21}$). Values of $j_{21}$ in the V-band are included in the table.

\begin{figure*}
    \begin{center}
        \includegraphics[width=\textwidth]{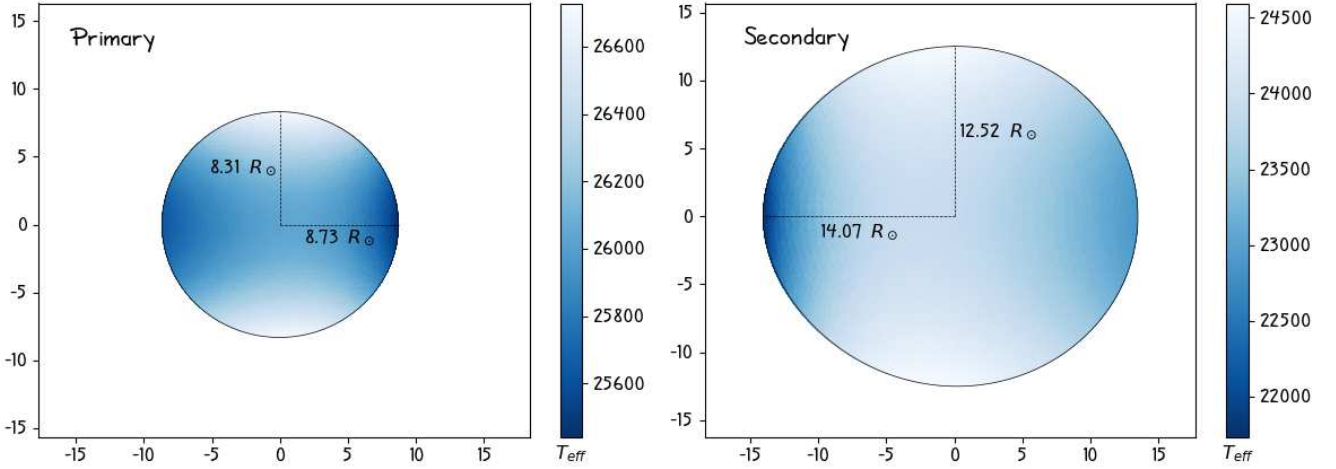} 
    \end{center}
\caption{ Surface temperature distribution and the shape of the primary (left) and secondary (right) components of the BLMC-01 system as seen from the side. The same scale is used in both panels. Polar and maximum (towards the mass center) radii are shown. }
\label{view}
\end{figure*}

As the stars are slightly deformed, the radii of the stars given in the table correspond to radii of spheres that have the same volume as the stars. As a measure of oblateness a value $(R_{max}-R_{min})/R_{max}$ is given. To help understanding the configuration, a ratio of polar star radius to polar Roche lobe radius ($R_{p}/R_{p,Roche}$) is also included.
In Fig.~\ref{view} we show the shape of the primary and secondary components of the BLMC-01 system as seen from the side (i.e. at quadrature and for $i=90^\circ$) and the temperature distribution over the projected surface.

We would like to point out that for partially eclipsing detached systems the sum of the radii is well defined, but different combinations of individual radii may give solutions that are similarly probably. This anti-correlation between the radii may significantly increase the uncertainty of their determination, as in the case of BLMC-02. The proximity effects however, are known to break this degeneracy, as they are a strong function of the star radius (relative to the components separation). Indeed, the deformation of the BLMC-01 secondary helped in an unambiguous determination of the radii of the components. This can be well seen comparing the uncertainties for both systems for which the data are of similar quality.

To calculate final rotational velocities ($v_{rot}$) from $v \sin i$ we assumed that the rotational axes are parallel to the orbital axis.

In our calculations and presentation of the results we make use of the nominal solar constants following the IAU 2015 Resolution B3 \citep{mamajek2015,prsa2016}.

\begin{table*}
\begin{center}
\caption{Physical parameters and other properties of the systems}
\begin{tabular}{|l|cc||cc|}
\cline{1-5}
  &  \multicolumn{2}{c||}{ BLMC-01}  &   \multicolumn{2}{c|}{ BLMC-02} \\
\hline
Parameter          & Primary & Secondary   & Primary & Secondary  \\
\hline\hline
mass [$M_\odot$]   & \VyE{\mPriBLMCuno}{\mPriBLMCunoErr}   & \VyE{\mSecBLMCuno}{\mSecBLMCunoErr}   &  \VyE{\mPriBLMCdos}{\mPriBLMCdosErr} &  \VyE{\mSecBLMCdos}{\mSecBLMCdosErr} \\ 
radius [$R_\odot$] & \VyE{\rPriBLMCuno}{\rPriBLMCunoErr}   & \VyE{\rSecBLMCuno}{\rSecBLMCunoErr}   &  \VyE{\rPriBLMCdos}{\rPriBLMCdosErr} &  \VyE{\rSecBLMCdos}{\rSecBLMCdosErr}  \\
$\log g$ [cgs]     & \VyE{\loggPriBLMCuno}{\loggPriBLMCunoErr}  &  \VyE{\loggSecBLMCuno}{\loggSecBLMCunoErr}   & \VyE{\loggPriBLMCdos}{\loggPriBLMCdosErr}  & \VyE{\loggSecBLMCdos}{\loggSecBLMCdosErr} \\ 
temperature [K]   & \VyE{\teffPriBLMCuno}{\teffPriBLMCunoErr}   & \VyE{\teffSecBLMCuno}{\teffSecBLMCunoErr}   &  ----- &  -----  \\ 
$\log L$ [$L_\odot$] & \VyE{\logLPriBLMCuno}{0.010}   & \VyE{\logLSecBLMCuno}{0.003}   & ----- &  -----  \\
$M_{bol}$ [mag]        & \VyE{\mbolPriBLMCuno}{\mbolPriBLMCunoErr}   & \VyE{\mbolSecBLMCuno}{\mbolSecBLMCunoErr}   & ----- &  -----  \\
$BC_{V}$ [mag]       & \VyE{\BCvPriBLMCuno}{\BCvPriBLMCunoErr}   & \VyE{\BCvSecBLMCuno}{\BCvSecBLMCunoErr}   & ----- &  -----  \\
$j_{21,V}$         & \multicolumn{2}{c||}{0.856 $\pm$ 0.002}  & \multicolumn{2}{c|}{0.985 $\pm$ 0.005} \\
$V$ [mag]            & \VyE{\mVPriBLMCuno}{\mVPriBLMCunoErr}   & \VyE{\mVSecBLMCuno}{\mVSecBLMCunoErr}   & \VyE{\mVPriBLMCdos}{\mVPriBLMCdosErr}  & \VyE{\mVSecBLMCdos}{\mVSecBLMCdosErr}  \\
$R$ [mag]            &  -----     & -----     & \VyE{\mRPriBLMCdos}{\mRPriBLMCdosErr}  & \VyE{\mRSecBLMCdos}{\mRSecBLMCdosErr}  \\
(V-$I_C$) [mag]      & \VyE{\VIPriBLMCuno}{\VIPriBLMCunoErr}   & \VyE{\VISecBLMCuno}{\VISecBLMCunoErr}   & \VyE{\VIPriBLMCdos}{\VIPriBLMCdosErr}  & \VyE{\VISecBLMCdos}{\VISecBLMCdosErr}  \\
(V-K) [mag]          & \VyE{\VKPriBLMCuno}{\VKPriBLMCunoErr}   & \VyE{\VKSecBLMCuno}{\VKSecBLMCunoErr}   & -----     & -----     \\
$v_{rot}$ [km/s]                   & \vrotBFPriBLMCuno & \vrotBFSecBLMCuno           &  \vrotBFPriBLMCdos & \vrotBFSecBLMCdos \\
oblateness                   & 0.05 &   0.11        &  0.04 & 0.04 \\
$R_{p}/R_{p,Roche}$          & 0.68 &   0.88        &  0.68 & 0.65 \\
E(B-V) [mag]               & \multicolumn{2}{c||}{ \VyE{\avgEbvBLMCuno}{\avgEbvBLMCunoErr} }         &  \multicolumn{2}{c|}{ \VyE{\avgEbvBLMCdos}{\avgEbvBLMCdosErr}} \\
orbital period [days]      & \multicolumn{2}{c||}{ \VyE{\perBLMCuno}{\perBLMCunoErr} }    &  \multicolumn{2}{c|}{ \VyE{\perBLMCdos}{\perBLMCdosErr} } \\
semi-major axis [$R_\odot$]& \multicolumn{2}{c||}{\VyE{\smaBLMCuno}{\smaBLMCunoErr}} & \multicolumn{2}{c|}{\VyE{\smaBLMCdos}{\smaBLMCdosErr}} \\
inclination                & \multicolumn{2}{c||}{ \VyE{\inclBLMCuno}{\inclBLMCunoErr}  }  &  \multicolumn{2}{c|}{ \VyE{\inclBLMCdos}{\inclBLMCdosErr} } \\
eccentricity               & \multicolumn{2}{c||}{ \VyE{\eccBLMCuno}{\eccBLMCunoErr} }    &  \multicolumn{2}{c|}{ \VyE{\eccBLMCdos}{\eccBLMCdosErr} } \\
$\omega$ [rad]              & \multicolumn{2}{c||}{ \VyE{\aopBLMCuno}{\aopBLMCunoErr} }    &  \multicolumn{2}{c|}{ \VyE{\aopBLMCdos}{\aopBLMCdosErr} } \\
$\dot{\omega}$ [rad/d]              & \multicolumn{2}{c||}{ \VyE{\daopBLMCuno}{\daopBLMCunoErr} }    &  \multicolumn{2}{c|}{ \VyE{\daopBLMCdos}{\daopBLMCdosErr} } \\
\hline
\end{tabular}
\end{center}
\tablecomments{ Color indices and passband magnitudes are observed (reddened) values.}
\label{properties}
\end{table*}

\subsection{Rotation}
During the binary evolution due to the mutual tidal influence the orbits of the components are slowly circularized and their rotational velocities synchronized.
Knowing the physical properties of the stars we can compare the rotational angular velocity of components with the synchronous ($\Omega_{sync}$) and pseudo-synchronous angular velocity ($\Omega_{ps}$) for each system, to check whether they are synchronized or not.

We obtain $\Omega_{sync} = 1.161$ radian per day for BLMC-01 and $\Omega_{sync} = 1.471 \, d^{-1}$ for BLMC-02, while using equations from \citet{hut:1981} we obtain $\Omega_{ps} = 1.162 \, d^{-1}$ and $\Omega_{ps} = 1.529 \, d^{-1}$ for BLMC-01 and BLMC-02, respectively.

Using radii of the stars from our models we can calculate the angular rotational velocity of the components of BLMC-01: $1.91 \pm 0.07 \, d^{-1}$ and $1.12 \pm 0.03  \, d^{-1}$, which means that the primary is still rotating super-synchronously, while the secondary is already synchronized, probably because of its larger radius resulting from its more advanced evolutionary state (it is more massive than the primary).

For BLMC-02 the angular velocities of the components are $1.51 \pm 0.04 \, d^{-1}$ and $1.66 \pm 0.05 \, d^{-1}$. This means that the larger and more massive primary is already rotating synchronously, while the smaller secondary has not yet synchronized its rotation.

\subsection{Distances}
Here we will present two approaches for the determination of the distance to a system in the LMC. One is a direct measurement using the bolometric flux scaling method, and the other one, where the distance is inferred from the assumed LMC distance. 

In the first one a distance modulus (DM) is calculated using the absolute V-band magnitude of a system ($M_{V}$) and the observed magnitude corrected for extinction determined in Section~\ref{subsec_extinction} ($V_{0} = V_{obs} - A_{V}$) using the formula: $ DM = V_{0} - M_{V}$. 
The value of $M_{V}$ is calculated using the derived physical parameters of the components and applying a bolometric correction (BC$_{V}$) as obtained for the observed colors and surface gravities from the calibration of \cite{worthey:2011}: $ M_{V} = M_{bol} - BC_{V} $.
The final equation is:

$$ DM_{bol} =  V_{obs} - M_{bol} + BC_{V} - A_{V} $$

Following this procedure for BLMC-01 we obtain a distance modulus DM$_{bol} = \BLMCunoDM{} \pm \BLMCunoDMerr{}$ mag.

For BLMC-02, using the measured reddening and transforming the resulting (V-I) color to temperature, we would obtain a distance modulus about $17.5$ mag, which is clearly wrong for a system that is expected to belong to the LMC. As described before, to obtain a consistent distance and using the measured reddening, a temperature of the primary around 34000-35000 K is needed.

An alternative way to reliably estimate the distance to both systems is to use the known distance to the LMC and apply the geometrical correction for the position of the system in the plane of the LMC (see Section~\ref{sec:data})
We used the LMC distance (DM$_{LMC}$) determined from the analysis of binary systems composed of late-type giants by \citealt{pietrzynski:2019}) and applied the geometrical correction (GC) based on the "PMs+Young" LMC model of \cite{vandermarel2014}:
$$DM_{gc} = DM_{LMC} + GC $$
Using this method we determined the expected distance moduli (DM$_{gc}$) to the systems to be \DMcgeomBLMCunoDepth mag for BLMC-01, and \DMcgeomBLMCdosDepth mag for BLMC-02. To account for the depth of the LMC, the quoted uncertainties include both the total error of the LMC distance (0.028 mag), and the scatter of 20 individual distance measurements (0.026 mag) used by \citet{pietrzynski:2019}. For BLMC-01, this distance is consistent within errors with the distance modulus determined with the first method (DM$_{bol}$).

\subsection{Surface Brightness -- Color relation}
To compare our results with the existing surface-brightness -- color relations we need (V-K) color and the surface brightness in V-band ($S_V$). As already mentioned, for BLMC-01 the K-band light curve is available covering both eclipses. From the light curve solution, the individual K-band magnitudes of each component were derived and then their infrared (V-K) colors were calculated.
Knowing the distance to this system, we can also calculate $S_V$ of the components. To do this we used the distance modulus DM$_{gc}$, which provides much better precision than the DM determined using the bolometric flux scaling method. In the future we are going to use the same methodology for the whole sample. The following equations were used to obtain the surface brightness:

$$ \theta[mas] = 1.337 \times 10^{-5} \times R [km] / D[pc] $$
$$ S_V = m_0 + 5 \, \log \, \theta [mas]$$

\begin{figure}
    \begin{center}
        \includegraphics[width=0.48\textwidth]{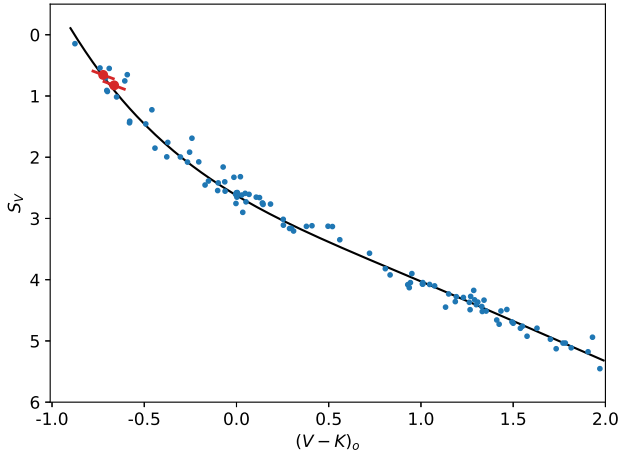} 
    \end{center}
\caption{Surface brightness - (V-K)--color relation. Interferometric measurements (blue points) and the fit (solid line) come from \citet{challouf:2014}. Red points are components of BLMC-01 with errors corresponding to the uncertainty of the reddening estimation.}
\label{SBrel}
\end{figure}

In Fig.~\ref{SBrel} we show the comparison of our measurements for the components of the BLMC-01 system  with the SBCR of \cite{challouf:2014}. Solid line is the fit of the relation for stars of all spectral types (blue points) given in Eq.~10 by these authors. Our measurements are consistent with their results. Once we have similar data for all 18 components of 9 systems from the sample, we will be able to improve the calibration of the blue part of this relation, which now suffers from a lack of precise measurements.

\subsection{Evolutionary status}
The evolution of early-type stars is still not well understood and a discrepancy between the observations and models are often found. One example is the mass discrepancy problem. It was first noticed by \cite{herrero:1992}, where the evolutionary masses were systematically higher than the ones determined with spectroscopy, and is still being raised by other authors \citep[e.g.][]{pavlovski:2018}. There are also not many systems composed of early-type components for which accurate physical parameters are known. This makes a proper calibration of theoretical models harder for this type of stars.

In this study we have obtained precise and accurate physical parameters for the components of the BLMC-01 system. Below we will test how well our measurements are described by modern evolutionary models. 

\begin{figure}
    \begin{center}
        \includegraphics[width=0.50\textwidth]{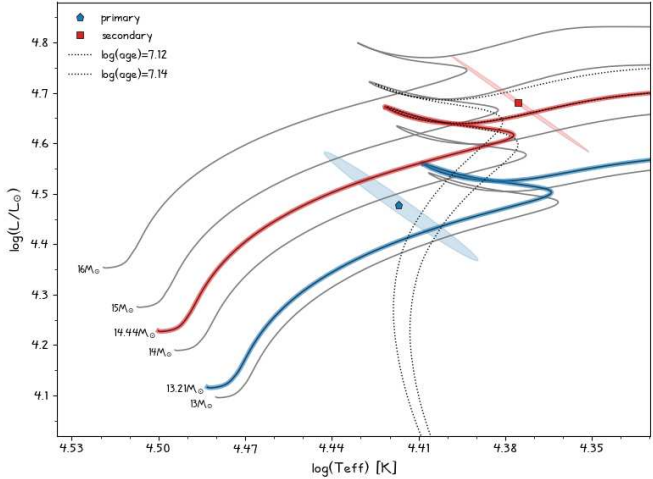} \\ 
    \end{center}
    \caption{HR diagram with the position of the components of BLMC-01 on a grid of evolutionary tracks of \citet{choi:2016} for masses 13 to 16 $M_\odot$ (thin gray lines). Two interpolated tracks for the masses of the primary and secondary are shown in color (thick lines). All tracks start at zero-age main sequence. Error ellipses are shown for the components. }
    \label{evol_choi}
\end{figure}

First we compared our results with the evolutionary tracks of \citet{choi:2016}. We used a set models for the LMC metallicity and masses from 13 to 16 $M_\odot$ and also two interpolated tracks for the masses of our components (see Fig.~\ref{evol_choi}). All these models were calculated with the overshooting value of $f_{ov,core}=0.016$, which is an equivalent of $\alpha=0.2$ in the step overshooting scheme.

The tracks are consistent within the errors with our results, but there are few points that have to be noted. First, these models suggest that the secondary has already left the main sequence, which would be interesting but because of the fast phase of evolution, would be a rare coincidence and would have to be confirmed independently.
Second, the uncertainty of the temperature determination is high but strongly correlated between the two components so any shift would move both systems in the same direction in the HR diagram. If we adjust the temperature of the secondary for it to lie exactly on the track, the primary would be overluminous, and if we do the same for the primary, the secondary would be underluminous.
The age of the system in this model is 13.8 million years.

To check the dependence on the model and the assumed parameters, we also used the grid of evolutionary tracks of \citet{brott:2011}. This grid covers masses from 5 to 60 $M_\odot$ and were calculated for three different metallicities (MW, LMC, and the Small Magellanic Cloud) and a wide range of surface rotation velocities (0-400 km/s). For all models the same overshooting was used ($\alpha=0.335$). We chose models with the LMC metallicity, initial velocity of $\sim$140~km/s, and interpolated them to a given mass when necessary.

\begin{figure}
    \begin{center}
        \includegraphics[width=0.50\textwidth]{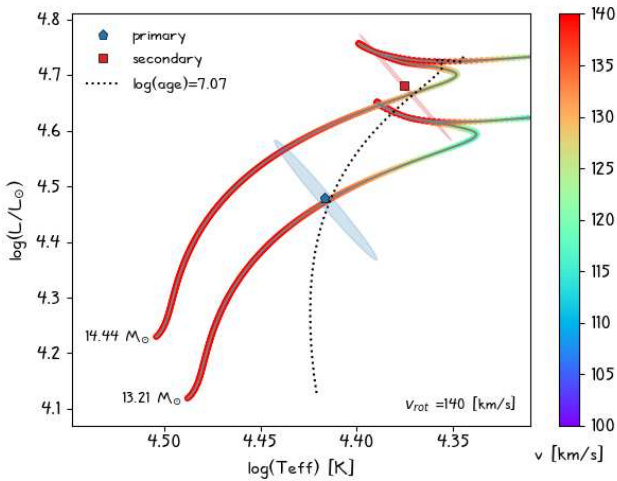} \\ 
    \end{center}
    \caption{Similar to Fig.~\ref{evol_choi}, but interpolated evolutionary tracks from \citet{brott:2011} are shown. Rotation velocity of the components along the evolutionary tracks is color-coded.}
    \label{evol_brott}
\end{figure}

As can be seen in Fig.~\ref{evol_brott} our results are consistent with these evolutionary  models and in this case the secondary is still on the main sequence, which is a more likely scenario. This is probably the effect of higher overshooting, which extends the main sequence towards lower temperatures. Contrary to the models of \citet{choi:2016} both components lie close to their corresponding evolutionary tracks and at roughly the same distance from them.
The model rotational velocities of the components are also consistent with our measurements.
In general we find the models of \citet{brott:2011} to describe better our data, which points to higher overshooting values for early-type stars.
The age of the system in this model is 11.7 million years.

Both presented grids were calculated for single star evolution, which should work reasonably well for our system. The stars are still small enough not to interact directly with each other, although the interaction through tidal forces has probably taken place -- the larger components have already synchronized their rotation and orbital velocities.
However, as the stars started their evolution on the main sequence being much smaller than they are now, for a significant part of their life tidal forces were negligible.

\section{Summary}\label{sec:summary}
Accurate physical parameters for early-type stars are still lacking, and the analysis of such stars in binary systems provide a very good means for the study of their properties and evolution.

In our project we have selected for observations and analysis  9  early-type well-detached, SB2, eclipsing binary systems in the Large Magellanic Cloud and here we present the results for first two of them.
Except for a rough determination of the spectral type neither of them was analyzed so far and their physical properties were not known before.

Using the high-quality data that we have acquired for these systems, we could measure their parameters with unprecedented accuracy and precision. For the masses it is: $0.6-1\%$, and for the radii: $0.4-3\%$ depending on the component. The determined masses are from 13 to 20 $M_\odot$ and the radii from 8 to 13 $R_\odot$, which means that the stars are now in the middle or at the end of their main-sequence evolution.

\citet{lebouquin:2017} reported problems with the radial velocity semi-amplitudes determined for massive stars, which could lead to erroneous masses of the components. They however used a different method of mass determination (combining  their astrometric orbits with double-lined radial velocity amplitudes) for which eclipses are not necessary to obtain the system inclination. For this reason in their sample many long-period systems with low inclinations are present. The derivation of RVs for such binary systems is however much harder due to lower amplitudes and severe blending of the profiles. The latter was indicated by these authors as the most probable cause of the problem. As in our case the amplitudes are very high and the profiles are not blended, we are confident that our measurements are free from this potential problem.

For one system, for which we could determine the temperatures we compared our results with evolutionary models and found them to be consistent within uncertainties. At least for this one system there is no strong evidence for mass discrepancy as reported by several authors before. However, we have to point out, that because of the difficulty in determination of the reddening, the error of the temperature determination is high and more advanced spectral analysis will be necessary to decrease the uncertainty. The same analysis will be necessary for the second system for which we could not determine a reliable temperature.

For one of the systems we have measured a distance modulus using the bolometric flux scaling method. The obtained value, \BLMCunoDM{}$\pm$\BLMCunoDMerr{} mag, is in good agreement with the recent LMC distance determination of \citet{pietrzynski:2019} corrected for the non-central position of the star in the galaxy (\DMcgeomBLMCuno mag).

The precision of the radius determination from our analysis is especially important for the main aim of our project to use these systems to calibrate the surface brightness -- color relation and to make early-type binary systems a precise tool for distance determination in the Local Group. Although it is not possible to calibrate this relation using results for only one system, we checked that the recent calibration of \cite{challouf:2014}, partly based on VEGA/CHARA interferometric measurements, is consistent with our measurements.

We are going to calibrate the SBCR once we finish the analysis of the rest of the systems and have 18 data points in the blue part of this relation.

\section{Acknowledgments}
MT acknowledges financial support from the Polish National Science Center grant PRELUDIUM 2016/21/N/ST9/03310. The research leading to these results has received funding from the European Research Council (ERC) under the European Union's Horizon 2020 research and innovation program (grant agreement No 695099). WG and GP also gratefully acknowledge financial support for this work from the BASAL Centro de Astrofisica y Tecnologias Afines (CATA, AFB-170002). WG acknowledges financial support from the Millennium Institute for Astrophysics (MAS) of the Iniciativa Milenio del Ministerio de Econom\'{i}a, Fomento y Turismo de Chile, project IC120009. BP acknowledges support for this work from the Polish National Science Center grant SONATA 2014/15/D/ST9/02248. RPK has been supported by the Munich Excellence Clusters Universe and Origins. Support from the Polish National Science Centre grants MAESTRO UMO-2017/26/A/ST9/00446 and from the IdPII 2015 0002 64 grant of the Polish Ministry of Science and Higher Education is also acknowledged. 

This research is based on observations collected at the European Southern Observatory under ESO programmes: 098.D-0263(A), 097.D-0.400(A), 0100.D-0339(A) and 0102.D-0469(A). The LMC background image used in Fig.~\ref{skyposition} was kindly provided by I. Soszy\'{n}ski. MT would like to thank J. Ostrowski for helpful discussions. We would also like to thank the anonymous referee for very constructive comments.

This research has made use of NASA's Astrophysics Data System Service.

\vspace{5mm}
\facilities{VLT:Kueyen (UVES), Magellan:Clay (MIKE)}

\software{
\texttt{ESO Reflex} \citep[][\url{http://www.eso.org/sci/software/esoreflex/}]{freudling:2013}, \\
\texttt{PHOEBE} \citep[][\url{http://phoebe-project.org/1.0}]{prsa:2005},\\
\texttt{RaveSpan} \citep[][\url{https://users.camk.edu.pl/pilecki/ravespan/index.php}]{pilecki:2017}, \\
\texttt{JKTEBOP} \citep[][\url{http://www.astro.keele.ac.uk/~jkt/codes/jktebop.html}]{southworth:2007},\\
}

\bibliographystyle{aasjournal}
\bibliography{taormina2019}

\end{document}